\documentclass[preprint,12pt]{aastex}           
\usepackage{natbib,graphics}

\newcommand{\be}{\begin{equation}}
\newcommand{\ee}{\end{equation}}

\newcommand{\bea}{\begin{eqnarray}}
\newcommand{\eea}{\end{eqnarray}}
\newcommand{\rc}{r_{cor}}
\newcommand{\ri}{r_{i}}
\newcommand{\ro}{r_{o}}
\newcommand{\rP}{r_{+}}
\newcommand{\rM}{r_{-}}

\shorttitle{Corotation Resonance}
\shortauthors{Silbergleit, Wagoner}

\begin{document}
\title{Corotation Resonance and Diskoseismology Modes\\
 of Black Hole Accretion Disks}

\author{Alexander S. Silbergleit\altaffilmark{1}}
\affil{Gravity Probe B, Stanford University, Stanford, CA 94305--4085}

\author{Robert V. Wagoner\altaffilmark{2}} 
\affil{Department of Physics and KIPAC, 
Stanford University, Stanford, CA 94305--4060}
 
\altaffiltext{1}{gleit@stanford.edu} 
\altaffiltext{2}{wagoner@stanford.edu}

\begin{abstract}

We demonstrate that the corotation resonance affects only some non--axisymmetric g--mode oscillations of thin accretion 
disks, since it is located within their capture zones. Using a more general (weaker radial WKB approximation) formulation of the governing equations, such g--modes, treated as perfect fluid perturbations, are shown to formally diverge at the position of the corotation resonance. 
A small amount of viscosity adds a small imaginary part to the eigenfrequency which has been shown to induce a secular instability (mode growth) if it acts hydrodynamically. The g--mode corotation resonance divergence disappears, but the mode magnitude can remain largest at the place of the corotation resonance. For the known g--modes with moderate values of the radial mode number and axial mode number (and any vertical mode number), the corotation resonance lies well outside their trapping region (and inside the innermost stable circular orbit), so the observationally relevant modes are unaffected by the resonance. The axisymmetric g--mode has been seen by Reynolds \& Miller in a recent inviscid hydrodynamic accretion disk global numerical simulation. We also point out that the g--mode eigenfrequencies approximately obey the harmonic relation $\sigma\propto m$ for axial mode numbers $|m|\geq 1$.

\end{abstract}


\section{Introduction}
In principle, all adiabatic perturbations of equilibrium models of accretion disks can be analyzed in terms of global normal modes. The pioneering studies of Shoji Kato and his group and the more recent work of our group have focused on accretion disks around black holes, so that no complications from boundary layers are involved. For a recent review of `relativistic diskoseismology', see \citet{k01}. A short summary of observationally relevant results from our recent analyses of the low-lying spectrum, which consists of g--modes \citep{per} (hereafter referred to as RD.I), c--modes \citep{swo} (hereafter referred to as RD.II), fundamental p--modes \citep{osw} (hereafter referred to as RD.III), and other p--modes \citep{rd4} (hereafter referred to as RD.IV), is given by \citet{wso}. 

Local analyses (restricted radial interval) have also played an important role in our understanding of these perturbations \citep{kfm}. Indeed, one of the motivations for this paper was the search by \citet{lgn} for a dynamical instability via such an analysis of traveling waves impinging on the corotation (and Lindblad) resonance. They argued that no modes were likely to be dynamically unstable, since amplification at the corotation resonance would not occur in realistic thin disks.

In section 2 we summarize the foundations (assumptions and equations) of our approach. In section 3 we examine the corotation resonance location and show that only certain non--axisymmetric g-modes may be affected by the resonance. In section 4 we investigate the behavior of the vertical and radial eigenfunctions near the corotation resonance, exhibiting the local and global divergence, and thus prove that the range of the eigenfrequencies of non--axisymmetric g-modes is reduced to its upper part (specified by the rotational frequency, radial epicyclic frequency, and the azimuthal wave number of the mode). In section 5 we discuss effects of introducing viscosity and buoyancy. We also comment on the possible observational relevance of the spectrum of the g--modes.

\section{Basic Assumptions and Equations}

We take $c=1$, and express all distances in units of $GM/c^2$ and all frequencies in units of $c^3/GM$ 
(where $M$ is the mass of the black hole) unless otherwise indicated. We employ the Kerr metric to study 
a thin accretion disk. The equilibrium disk is taken to be described by the standard relativistic thin disk model \citep{nt,pt}. The velocity components $v^r=v^z=0$, and the disk semi-thickness $h(r)\sim c_s(r,0)/\Omega\ll r$, 
where $c_s(r,z)$ is the speed of sound. The key frequencies, associated with free-particle orbits, are 
\bea \label{3freq}
\Omega(r) & = & (r^{3/2}+a)^{-1}\; , \nonumber \\
\Omega_\perp(r) & = & \Omega(r)\left(1-4a/r^{3/2}+3a^2/r^2\right)^{1/2}\; , \nonumber \\
\kappa(r) & = & \Omega(r)\left(1-6/r+8a/r^{3/2}-3a^2/r^2\right)^{1/2}\; 
\eea
the rotational, vertical epicyclic, and radial epicyclic frequencies, respectively. The angular momentum parameter 
$a=cJ/GM^2$ is less than unity in absolute value. 
The inner edge of the disk is at approximately the radius of the last stable free-particle circular orbit $r=r_i(a)$, 
where the epicyclic frequency $\kappa(r_i)=0$. So all the relations we use are for $r>r_i$, where $\kappa(r)>0$.
The outer disk radius is denoted by $r_{o}$. 

We apply the general relativistic formalism that \citet{il} developed for perturbations of purely rotating perfect fluids, although the effects of viscosity are included in the equilibrium model (and some of our notation differs from theirs, see below). The pressure $p$ is much less than the mass--energy density $\rho$. We neglect the self-gravity of the disk, which is a good approximation since the ratio of disk to black hole mass is usually very small (see RD.IV, section 7). \citet{il} then show that one can express the Eulerian perturbations of all physical quantities through a single function 
\be
\delta V= \frac{\delta p}{\rho\beta\omega}\; . \label{deltaV}
\ee
Due to the stationary and axisymmetric equilibrium, the angular and time dependences are factored out as 
$\delta V = V(r,z)\exp[i(m\phi + \sigma t)]$), and the master equation (39) of \citet{il} for the function $V$ [see also RD.I, equation (2.21)] assumes the form 
\be
\frac{1}{r}\frac{\partial}{\partial r}\left[r^2\Upsilon^3\rho\left(\frac{\omega^2}{\omega^2-\kappa^2}\right)
\frac{\partial V}{\partial r}\right] +
\frac{\partial}{\partial z}\left[r\Upsilon\rho\left(\frac{\omega^2}{\omega^2-N_z^2}\right)\frac{\partial V}{\partial z}\right] + r\Upsilon\beta\omega\Phi V = 0 \; .
\label{basicPDE}
\ee
The corotation frequency $\omega$ is related to the eigenfrequency $\sigma$ by
\be
\omega(r,\sigma) = \sigma+m\Omega(r) \; . \label{omega}
\ee
The buoyancy frequency is dominated by its vertical component $N_z$ (RD.I). The Kerr metric component $g^{rr}\equiv\Upsilon^2(r)$ and four-velocity component $u^0=dt/d\tau\equiv\beta(r)$ are both of order unity (in Boyer--Lindquist coordinates); their expressions can be found in RD.I. Compared with \citet{il}, the definitions of $\sigma$ and $\omega$ are switched, while their $\gamma$ becomes our $\beta$. 
The function $\Phi(r,z,\omega,m)$ depends on various properties of the unperturbed disk, as well as the mode eigenfrequency, $\sigma$, and its azimuthal wave number, $m$. 

To simplify the analysis, we consider barotropic disks [$p=p(\rho)$], so the buoyancy frequency $N_z=0$. Since this frequency should be lower than other characteristic frequencies, and in general vanishes on the mid--plane of symmetry $z=0$, this should be a good approximation except possibly in the neighborhood of the corotation resonance [where $\omega(r)=0$]. Hydrostatic equilibrium then provides the following unperturbed density and pressure distributions ($\Gamma>1$ is the adiabatic index):
\be
\rho(r,y)=\rho_0(r)(1-y^2)^{g}\; ,\quad p(r,y)=p_0(r)(1-y^2)^{g+1}\;, \quad 
g\equiv1/(\Gamma-1) >0 \; ; \label{dens/pres}
\ee
the disk surfaces are at $y=\pm1$, with the new coordinate $y$ related to the vertical coordinate $z$ through the characteristic disk semi-thickness $h(r)$:
\be 
y={z\over h(r)}\,\sqrt{\Gamma-1\over2\Gamma}, \qquad
h(r) = \frac{1}{\beta(r)\Omega_\perp(r)}\,\sqrt{\frac{p_0(r)}{\rho_0(r)}}\; . \label{y}
\ee
The speed of sound $c_s(r,y)$ is specified by 
\be
c_s^2 =\Gamma\,{p}/{\rho}= \Gamma(h\beta\Omega_\perp)^2(1-y^2) \; .
\label{csh}
\ee

With these simplifications, for the function $\Phi$ which appears at the end of the master equation (\ref{basicPDE}) we  obtain: 
\be
\Phi(r,z) = \frac{\rho\beta\omega}{c_s^2}\; + \label{Phi}
\ee
\[
\frac{1}{\Upsilon r^2}\,\frac{\partial}{\partial r}
\left[\frac{\rho\Upsilon^3 r^2}{\beta^2(\omega^2-\kappa^2)}\,\frac{\partial}{\partial r}(\beta\omega)
- \frac{2\rho\Upsilon r\Omega^z\,(m+\beta\omega u_\phi)}{\beta^3(\omega^2-\kappa^2)}
\right]
- \frac{\rho\omega(m+\beta\omega u_\phi)^2}{\beta^3\Upsilon^2 r^2(\omega^2-\kappa^2)}\; .
\] 
The quantity $\Omega^z$ is a component of the angular velocity four--vector, while $u_\phi$ is another component of the fluid four--velocity. Like all other quantities in the expression (\ref{Phi}) except $\rho$ and $c_s$, they are functions of $r$ alone.

Because of the structure of the density and speed of sound as functions of the coordinates, the equation (\ref{basicPDE}) does not allow for an exact separation of variables in either of their two pairs, $\{r,z\}$ or $\{r,y\}$. Therefore in the past we adopted the (usually realistic) assumption of strong variation of the modes in the radial direction (characteristic radial wavelength $\lambda_r \ll r$), and used the asymptotic separation of variables based on it. In particular, we look for a separated solution to the master equation (\ref{basicPDE}) of the form 
\be
V = V_r(r)V_z(z,r)
\label{separ}  
\ee
Unlike the previous analyses, here we do not neglect the derivatives of functions of $\omega(r)$ [in addition to $V_r$ and $(\omega^2-\kappa^2)^{-1}$], since it varies much more strongly than $V_z$ and all other quantities near the corotation resonance (for $m\not=0$). However, we take all other functions out of the radial derivatives. Then, with the separation (\ref{separ}) also used, the master equation assumes the form 
\be
\frac{\Upsilon^2}{V_r}\frac{\partial}{\partial r}\left[\left(\frac{\omega^2}{\omega^2-\kappa^2}\right)\frac{\partial V_r}{\partial r}\right] + G(r)
= -\frac{1}{(1-y^2)^g V_z}\frac{\partial}{\partial z}\left[(1-y^2)^g\frac{\partial V_z}{\partial z}\right]
- \frac{\omega_*^2}{\Gamma h^2(1-y^2)}  \; . \label{WKBmaster}
\ee
We have introduced $\omega_*(r)\equiv\omega/\Omega_\perp$, and the function $G(r)$ is defined by
\be
G(r) \equiv G(r, a,\sigma, m) \equiv \frac{\beta\omega\Phi}{\rho} - \left(\frac{\beta\omega}{c_s}\right)^2 
= \label{G}
\ee
\[
\Upsilon^2\omega\frac{d}{dr}\left[\left(\frac{1}{\omega^2-\kappa^2}\right)\frac{d\omega}{dr}\right]
- \frac{2\Omega^z\omega}{\beta^2 r}\frac{d}{dr}\left(\frac{m+\beta u_\phi \omega}{\omega^2-\kappa^2}\right)
- \frac{(m+\beta u_\phi \omega)^2\omega^2}{\beta^2 r^2\Upsilon^2(\omega^2-\kappa^2)} \; .
\]
The left--hand side of equation (\ref{WKBmaster}) is a rapidly varying function of $r$, while the right--hand side is a rapidly varying function of $z$. Therefore within this (weak) radial WKB approximation both are equal to a slowly varying separation `constant' $S(r)$, which (in keeping with our previous convention) we denote instead by 
\be
S(r) \equiv [\Psi(r)-\omega_*^2]/(\Gamma h^2) \; .
\label{S&Psi}   
\ee
The vertical eigenvalue $\Psi$ is thus the redefined separation `constant'. Employing everywhere the new vertical coordinate $y$ defined by equation (\ref{y}), this radial WKB approximation then does indeed produce separated equations for $V_y = V_z(y,r)$ (a slowly varying function of $r$) and $V_r(r)$:
\be
\left(1-y^2\right)\frac{d^{\,2} V_y}{dy^2} - 2gy\,\frac{dV_y}{dy} + 
2g\left[\omega_*^2\, y^2+ \Psi\left(1-y^2\right)\right]V_y = 0 \; , \label{vert}
\ee
\be
\frac{d}{dr}\left[\left(\frac{\omega^2}{\omega^2 
- \kappa^2}\right)\frac{dV_r}{dr}\right] 
+ \left(G-\frac{\Psi-\omega_*^2}{\Gamma h^2}\right)\frac{V_r}{\Upsilon^2} = 0 \; . \label{rad}
\ee 
Away from the corotation resonance, $G\sim 1/r^2 + m^2/r^2$, which is seen to be of order $(h/r)^2$ smaller than its competing term $\omega_*^2/(\Gamma h^2) = [\beta\omega/c_s(r,0)]^2$ in equations (\ref{WKBmaster}) and (\ref{rad}). That is why $G(r)$ has been neglected in RD.I-IV.

Together with the proper boundary conditions, discussed in detail in the referenced papers, these two equations specify (slowly varying) vertical eigenvalues $\Psi=\Psi(\sigma,r)$ and eigenfrequencies $\sigma$, as well as the corresponding vertical ($V_y$) and radial ($V_r$) eigenfunctions for modes of all types. Lagrangian displacements are related to $V$ by the expressions
\bea
\xi^r \cong \frac{\omega\Upsilon^2}{\beta(\omega^2-\kappa^2)}\,
\frac{\partial V}{\partial r} \; , \qquad\quad 
\xi^z \cong\frac{1}{\beta\omega}\,\frac{\partial V}{\partial z}\; ,\nonumber\\
\xi^\phi \cong \frac{\beta^2[1-(2/r)(1-a\Omega)]}{i\omega}\,
\left[
\frac{d\Omega}{dr} + \frac{r\omega^z}{\beta^2(r^2 - 2r + a^2)}
\right]\,\xi^r\; .
\label {xirz}
\eea
[See equations (2.25) and (2.26) of RD.I.] Here $\omega^z$ is a component of the vorticity four-vector \citep{il}.

\section{The Corotation Resonance and Its Implications for Diskoseismology Modes}

\subsection{Frequency range and location of the corotation resonance}

As always, in view of the $\{\sigma\to-\sigma,\; m\to-m\}$ symmetry, we can restrict our consideration to $m\geq0$. For axially symmetric modes $m=0$ and $\omega\equiv\sigma\not=0$. However, for $m>0$ there might exist a point, $\rc=\rc(\sigma,m,a)$, where the corotation frequency $\omega$ goes to zero,
\be
\omega(\rc)=\sigma+m\Omega(\rc)=0 \; . \label{cordef}
\ee
This specifies the location of the corotation resonance for a given eigenfrequency. Since $\Omega(r)$ is a decreasing function of the radius, the unique solution $\rc=\rc(a,\sigma,m)$ to equation (\ref{cordef}) exists within the disk, $\ri<\rc<\ro$, for all eigenfrequencies in the range 
\be
m\Omega(\ro)<-\sigma<m\Omega(\ri)\; .\label{corrange}
\ee

It is important to compare $\rc$ with the Lindblad resonances at the radii $r_\pm=r_\pm(a,\sigma,m)$ ($\rM\leq\rP$) defined as the roots of 
$\omega^2(r)-\kappa^2(r)=0$. As pointed out in RD.I (note especially fig.~3), for $m>0$ they exist 
in a  wider range of the eigenfrequencies than (\ref{corrange}), namely, when
\be
m\Omega(\ro)<-\sigma<\max_{r_i<r<r_o}[m\Omega(r)+\kappa(r)]\; . \label{Lindrange}
\ee
This is also the maximum possible frequency range of g--modes, according to their definition specifying their capture zone as the interval between the Lindblad resonances where $\kappa^2(r) - \omega^2(r)>0$. 

So, whenever the corotation resonance exists, the Lindblad resonances are also present (but not necessarily vice versa). Moreover, since $\kappa(r)$ is positive inside the disk, we can write
$$
m\Omega(\rM)=-\sigma+\kappa(\rM)\;\;>\;\;-\sigma=m\Omega(\rc)\;\;>\;\; 
-\sigma-\kappa(\rP)=m\Omega(\rP)\;   
$$
Note that the inequalities here are based on the expressions involving $\sigma$. By looking at their counterparts with the angular velocity $\Omega(r)$ and invoking again its monotonic decrease, we conclude that
\be
\rM<\rc<\rP\; . \label{resord}
\ee
Hence, whenever the corotation resonance occurs, it lies between the Lindblad resonances.

\subsection{Corotation resonance and the diskoseismology modes}

The result (\ref{resord}) shows that the corotation resonance does not significantly affect both c--modes, captured in a region $\ri<r<r_c<\rM$ of the inner disk region (see RD.II), and inner and outer p--modes, residing respectively in $\ri<r<\rM$ and $\rP<r<\ro$ (see RD.III, IV). The only disk oscillations which could be strongly affected by it are thus the (non--axisymmetric) g--modes, since the region $\rM<r<\rP$ is their capture domain. So, the first significant result of the paper allows us to discuss only g--modes in the sequel. It should also be noted that g--modes are the most robustly determined, since their capture zone does not include either of the disk boundaries, where the physical conditions and validity of our assumptions are more uncertain.

Furthermore, for the effect of the corotation resonance to show up, the resonance itself must be present; this means that only those g--modes can be influenced by it whose eignefrequencies are in the corotation resonance range (\ref{corrange}).  All the g--mode eigenfrequencies found so far (see RD.I, Tables 1---3) belong, in fact, to the more restricted range
\be
m\Omega(\ri)<-\sigma<\max_{r_i<r<r_o}[m\Omega(r)+\kappa(r)]\; ,\label{complrange}
\ee
i.e., to the upper part of the maximum g--mode range (\ref{Lindrange}). The next section shows that this is not unexpected, since the g--modes with the eignefrequencies in the range (\ref{corrange}) are simply absent.

\section{Corotation Resonance Singularity and\\
the Frequency Range of Non--Axisymmetric g--Modes }

\subsection{Possible g--mode divergence at the corotation resonance}

Let us find out what can happen to those $m>0$ g--modes which have the corotation resonance in their 
capture domain, that is, eigenfrequencies in the interval (\ref{corrange}). The behavior of the eigenfunction, $V$, and its radial derivative, $\partial V/\partial r$, at the corotation resonance  needs to be examined, because one or both of them may become singular there. This would imply, in the first place, the singularity of the Lagrangian displacements by the formulas (\ref{xirz}), which is hardly acceptable for small perturbations. However, as shown in section 4.3, this is exactly what happens. Unfortunately, an integral divergence turns out also to be involved. 

To see this, one multiplies the master equation (\ref{basicPDE}) by $rV(r,z)$, integrates the result over the disk, and then integrates by parts taking into account that the unperturbed density ($\rho$) vanishes at the disk boundary, to obtain:
\be\label{ident}
\int\limits_{disk}\,r^2\Upsilon\beta\omega\Phi\,V^2\,drdz=
\int\limits_{disk}\,r^2\Upsilon\rho
\left[
\left(\frac{\partial V}{\partial z}\right)^2+
\frac{\Upsilon^2\omega^2}{\omega^2-\kappa^2}\,\left(\frac{\partial V}{\partial r}\right)^2
\right]\,drdz\; .
\ee
The left hand side of this identity must be finite for any reasonable perturbation, since it represents the relevant norm of it. The two sources of divergences on the right are: (a) the corotation resonance, and (b) the Lindblad resonances. 

For the source (b), the combination $(\omega^2-\kappa^2)^{-1}(\partial V/\partial r)$ has been shown to be finite at $r_\pm$ in RD.I and RD.III, and this conclusion holds with some new singular terms in the coefficient $G(r)$: by its definition (\ref{G}), $G(r)={\cal O}\left((\omega^2-\kappa^2)^{-2}\right)$ when $r\to r_{\pm}$.  Therefore the second term on the right of the identity (\ref{ident}) is finite (in fact, vanishes) at the Lindblad resonances. Moreover, $V$ and $(\omega^2-\kappa^2)^{-1}V$ are both finite there as well, so that there is no Lindblad resonance divergence in either the first term  on the right of equation (\ref{ident}), or in the term on the left stemming from the singularity in $G(r)$. 

However, it turns out that at the corotation resonance, the l.h.s.~remains finite but both terms on the r.h.s.~of the identity (\ref{ident}) generically diverge. To show this, we start with the first term, and transform it using the WKB separation, redefinition (\ref{y}) of the vertical variable, and expression (\ref{dens/pres}) for the density:
\[
\int\limits_{disk}\,r^2\Upsilon\rho\,\left(\frac{\partial V}{\partial z}\right)^2\,drdz=
\int_{\ri}^{\ro}\,\frac{r^2\Upsilon\rho_0}{h}\,\sqrt{\frac{\Gamma-1}{2\Gamma}}\,V_r^2\,dr
\int_{-1}^{1}\left(1-y^2\right)^{g}\left(\frac{\partial V_y}{\partial y}\right)^2dy=
\]
\be\label{divterm}
k\,\int_{\ri}^{\ro}\,{V_r^2I_y}\,dr \; .
\ee
Here $k>0$ is the proper average value of the positive function 
$r^2\Upsilon\rho_0 h^{-1}\sqrt{(\Gamma-1)/2\Gamma}$ of the radius, and we denoted
\be\label{Iy}
I_y=I_y(r)\equiv
\int_{-1}^{1}\left(1-y^2\right)^{g}\left[V_y^{'}(r,y)\right]^2dy \; . 
\ee
(Prime will now denote the derivative with respect to $y$.) Clearly, $I_y(r)\geq0$ for any $r$; if $I_y(r^*)=0$ for some $r=r^*$, then equation (\ref{Iy}) implies $V_y^{'}(r^*,y)\equiv 0$, so the vertical eigenfuntion $V_y(r^*,y)=\mbox{const}$. However, a constant never satisfies the vertical equation (\ref{vert}) except for the case $r=\rc$, $\Psi(\rc)=0$. The last condition is never true for g--modes; the demonstration requires nevertheless a slightly more sophisticated argument (following immediately).

\subsection{Vertical eigenvalue problem near the corotation resonance}

We rewrite the vertical equation (\ref{vert}) in the self-adjoint form,
\be
\left[
\left(1-y^2\right)^{g}V_y^{'}
\right]^{'} +
2g\left[
\Psi\left(1-y^2\right)+\frac{\omega^2}{\Omega_\perp^2}y^2
\right]
\left(1-y^2\right)^{g-1}V_y = 0 \; ,
\label{verts-a}
\ee
multiply by $V_y$ and integrate over the interval $(-1,\,1)$. Using integration by parts and the fact that $V_y(\pm1)$ must be finite, we thus obtain
\[
I_y(r)=\int_{-1}^{1}\left(1-y^2\right)^{g}\left(V_y^{'}\right)^2dy=
2g
\left[
\frac{\omega^2}{\Omega_\perp^2}\,\int_{-1}^{1}\,y^2\left(1-y^2\right)^{g-1}V_y^2dy + 
\Psi\,\int_{-1}^{1}\left(1-y^2\right)^{g}V_y^2dy
\right]
\; . 
\]
At the corotation resonance this equality becomes
\be\label{notzero}
I_y(\rc)=2g\Psi(\rc)\,\int_{-1}^{1}\left(1-y^2\right)^{g}\left[V_y(\rc,y)\right]^2dy\not=0
\; , 
\ee
unless $\Psi(\rc)=0$.  However, if this is the case, we can expand, in the vicinity of $\rc$, the eigenvalue and eigenfunction in the small parameter $\omega_*^2\equiv\omega^2/\Omega_\perp^2$ as
\[
{\Psi}=\Psi_1\,{\omega_*^2}+\ldots,\qquad V_y=V^{(0)}+V^{(1)}\,{\omega_*^2}+\ldots\; ,
\]
obtaining thus from the equation (\ref{verts-a}):
\[
{\cal L}V^{(0)}=0,\;\; 
{\cal L}V^{(1)}=-2g\,
\left[
1 - \Psi_1\left(1-y^2\right)
\right]
\left(1-y^2\right)^{g-1}V^{(0)}\ldots;\qquad 
{\cal L}v\equiv \left[
\left(1-y^2\right)^{g}v^{'}
\right]^{'}\; . 
\]
The only zero--order solution bounded together with its derivative at $y=\pm1$ is, of course, $V^{(0)}(y)\equiv 1$, and the solvability criterion of the problem for the first correction,
\[
0=\int_{-1}^{1}V^{(0)}{\cal L}V^{(1)}\,dy=-2g\,\int_{-1}^{1}\left[
1 - \Psi_1\left(1-y^2\right)
\right]
\left(1-y^2\right)^{g-1}\,dy\; ,
\]
provides $\Psi_1=-(\Gamma-1)/2$ by means of an easy calculation.
\footnote{The same derivation of this result was carried out in RD.III, section 3, for $m=0,\;\omega_*=\sigma/\Omega_\perp$; which specifics play, in fact, no role in it.} Hence
\be 
{\Psi(r)}/{\omega_*^2(r)}=- (\Gamma-1)/{2}+ {\cal O}\left(\omega_*^2(r)\right)=
- (\Gamma-1)/{2}+ {\cal O}\left((r-\rc)^2\right)<0,
\quad r\;{\rm near}\;\rc\; ,
\label{Psiatcor} 
\ee
which, by definition, corresponds to some p--mode, and not a g--mode (characterized by $\Psi/\omega_*^2>1$). Note that the last term in the equalities (\ref{Psiatcor}) holds due to the fact that $\omega$, and hence $\omega_*$, is to lowest order linear near $r=\rc$:
\be\label{omatcr}
\omega (r)=m\frac{d\Omega(\rc)}{dr}(r-\rc)+{\cal O}\left((r-\rc)^2\right) \; .
\ee

So $\Psi(\rc)\not=0$ for any g--mode, and therefore $I_y(\rc)\not=0$. Thus, from equation (\ref{divterm}), the question of whether the first term on the right of (\ref{ident}) diverges or not at the corotation resonance reduces to the same question about the norm of the radial eigenfunction (whose square we denote $I_r$):
\be\label{Ir}
\int\limits_{disk}\,r^2\Upsilon\rho\,\left(\frac{\partial V}{\partial z}\right)^2\,drdz\propto I_r
,\qquad
I_r\equiv \,\int_{\ri}^{\ro}\,{V_r^2}\,dr\; .
\ee
To obtain the answer, it remains only to investigate the behavior of $V_r(r)$ near $r=\rc$.

\subsection{Behavior of the radial eigenfunction near the corotation resonance}

Using the Taylor expansion (\ref{omatcr}), we see that the radial equation (\ref{rad}) near $\rc$ can be written as
\be
\frac{d^2 V_r}{dr^2} + \left[\frac{2}{r-\rc}+{\cal O}\left(1\right)\right]\,\frac{dV_r}{dr} + \left[\frac{Q_c^2}{(r-\rc)^2}+{\cal O}\left((r-\rc)^{-1}\right)\right]\,V_r = 0 \; , 
\label{radatcr}
\ee 
where 
\be
Q_c^2=\frac{\beta^2\kappa^2\Omega_\perp^2\Psi}{m^2c_s^2(r,0)}\,\left(\frac{d\Omega}{dr}\right)^{-2}\Biggl|_{r=\rc}\; ,
\label{not}
\ee
and all the higher order terms are integer powers of $(r-\rc)$. According to the analytical theory of second order ODEs [e.~g., \citet{olv}], the general solution to equation (\ref{radatcr}) near $\rc$ has the form:
\be
V_r(r)=C_+(r-\rc)^{\nu_+}(1+\ldots)+C_-(r-\rc)^{\nu_-}(1+\ldots)\; .
\label{wsol}
\ee
Here $C_{\pm}$ are some constants, and $\nu_\pm$ are the roots of the characteristic equation
\be
\nu^2+\nu+Q_c^2=0\; , \qquad \nu_{\pm}=-\frac{1}{2}\pm\sqrt{\frac{1}{4}-Q_c^2}\; .
\label{nu}
\ee
Note that for an isothermal thin disk model, \citet{lgn} have found a different behavior of traveling waves.

The results (\ref{wsol}) and (\ref{nu}) show that in the vicinity of the corotation resonance
\be
V_r(r)={\cal O}\left((r-\rc)^{-1/2}\right), \quad 
dV_r(r)/dr={\cal O}\left((r-\rc)^{-3/2}\right),\qquad r\to\rc \;. 
\label{watcr1}
\ee
The singularity is, in fact, even stronger if $Q_c^2<1/4$, but this is not important here. It is also rather an exceptional case: as shown in RD.I, for any axial mode number $m\geq1$ and radial mode number $n\geq0$, the vertical eigenvalue $\Psi=\Psi_{j}(r)$ grows indefinitely with the vertical mode number $j$ for all relevant radii, including $r=\rc$. So, by the formula (\ref{not}), $Q_c^2\geq 1/4$ for an infinite set of vertical eigenvalues, with the opposite inequality valid perhaps just for a few of them, if at all. Finally, the only case of a singularity weaker than the one in equations (\ref{watcr1}) is $Q_c^2<1/4$ and $C_-=0$ in expression (\ref{wsol}). However, this requires fine tuning of parameters. Therefore the estimates (\ref{watcr1}) always hold, as well as
\be
V_r^2={\cal O}\left((r-\rc)^{-1}\right), \quad 
\omega^2\,\left(dV_r/dr\right)^2={\cal O}\left((r-\rc)^{-1}\right),\qquad r\to\rc \; 
\label{watcr2}
\ee
implied by them.

\subsection{Local and global corotation resonance divergences.\\
Range of non--axisymmetric g--modes }

Formulas (\ref{watcr1}) combined with the expressions (\ref{xirz}) show that all the three components of the Lagrangian displacement are singular at the corotation resonance:
\[ \xi^r\sim (r-\rc)^{-1/2},\qquad \xi^z,\;\xi^\phi\sim (r-\rc)^{-3/2}\; . \]
However, note from equation (\ref{deltaV}) that the perturbation $\delta p/\rho \propto (r-\rc)^{1/2}$ vanishes there.

In addition to this local singularity of the physical quantities, a global one also exists. According to the first of the equalities (\ref{watcr2}), the integral $I_r$ defined in (\ref{Ir}) always diverges, hence so does the first term in the expression (\ref{ident}) for the norm of the eigenfunction. The second term there, in fact, also diverges via the second of the equalities (\ref{watcr2}). Such a combination of local and global perturbation singularities is definitely unacceptable. (The norm of an eigenfunction belonging to the discrete spectrum must be finite.)

We have thus proved that non-axisymmetric g--modes with eigenfrequencies in the corotation resonance range (\ref{corrange}) cannot exist within the framework of inviscid perturbations, so their actual frequency range (\ref{complrange}),
\be
m\Omega(\ri)<-\sigma<\max_{r_i<r<r_o}[m\Omega(r)+\kappa(r)]\equiv\kappa(r_m)+m\Omega(r_m),\qquad m=1,2,\ldots\; ,
\label{grange}
\ee
is the upper part of the maximum possible range (\ref{Lindrange}), for which the corotation resonance is absent within the disk. In contrast with this, the range of the axially symmetric g--modes, 
\be
\kappa(\ro)<|\sigma|<\max_{r_i<r<r_o}\kappa(r)\equiv\kappa_0(a)
\label{m=0range}
\ee 
includes low frequencies as well.

\section{Discussion and Conclusions}

\subsection{Non--axisymmetric g--modes with moderate radial and azimuthal wave numbers}

The established corotation resonance divergence could cast a shadow of doubt on the known results for non-axisymmetric g--modes, in particular, on those found in RD.I, especially since the WKB technique used there [see also \citet{rd4}] to calculate  eigenfrequencies is not fully sensitive to the presence of the corotation resonance. However, as pointed out in section 3.2, all the found eigenfrequencies belong to the proper range (\ref{grange}), $|\sigma_{mnj}|>m\Omega(\ri)$ ($m$, $n$, and $j$ are the azimuthal, radial and vertical mode numbers). That means that the corotation resonance lies inside the inner radius of the disk (where the gas is spiraling into the black hole). We now indicate why the g--modes (with moderate values of $m$ and $n$) are the most robust modes, and therefore astrophysically the most relevant.

For any azimuthal mode number $m$, we denote the largest eigenfequency $|\sigma_{mnj}|$ by $\sigma_m\equiv\kappa(r_m)+m\Omega(r_m)$ . In the limit $j\to\infty$, the trapping zone $\Delta r=r_+ - r_-\rightarrow 0$, with $r_+ > r_m >r_-$. We now consider moderate values of $m$ and $n$, but any value of the vertical mode number $j$. Then from equation (5.3) and Tables 1--3 of RD.I, it is seen that these g--modes occupy a very small frequency range below their maximum:
\[ \sigma_m-|\sigma_{mnj}|\lesssim (\kappa c_s/r\Omega)_{r_m} \; . \] 
Thus, as also seen from Fig.~3 and Fig.~5 of RD.I, it follows that 
\be
|\sigma_{mnj}|\approx m\Omega(r_i) \label{spectrum}
\ee  
for moderate values of $n$ and $m\geq 1$, with the approximate equality becoming quickly more and more exact as $m$ grows. Therefore for a given black hole, the largest frequency splitting should be due to the azimuthal mode number $m=0,1,2,\dots$, with the eigenfrequencies $|\sigma|\propto m$ for $m\geq 1$. We also note that with increasing values of $m$, the mode location $r_m\rightarrow r_i$ and its extent $\Delta r\rightarrow 0$. Then the mode leakage into the flow onto the black hole and the uncertainties in the physical conditions near $r_i$ become more important while the fractional modulation of the luminosity decreases. Therefore the low $m$ modes should be the most robust and observable. 

\subsection{QPO features: $3/2$ and other integer frequency ratios}

In this connection, one should note the observational claims that some of the quasi-stable high frequency QPOs in black hole binary X-ray sources have frequency ratios close to $3/2$ \citep{mr,rm}. This can be, in principle, explained by excitation of two (groups of) g--modes with $m=2$ and $m=3$, as suggested in RD.IV. 

Any such explanation for any QPO frequency implies a relation between the mass and angular momentum of the black hole in the corresponding X--ray binary. Indeed, using the general relation for the dimensional frequency $f = 3.23\times10^4(M_\odot/M)|\sigma|$ Hz, from formula (\ref{spectrum}) one obtains 
\be
M/M_\odot \cong 3.23\times10^4 m\Omega(r_i)/f_m(\mbox{Hz})\quad , \; m\ge 1 \; .\label{M-J}
\ee
Since $r_i=r_i(a)$, $\Omega(r_i)=(r_i^{3/2}+a)^{-1}\equiv{\cal F}(a)$ is a universal function of black hole angular momentum only. It is a monotonically increasing function, with ${\cal F}(-1)=0.038$ (extreme counter-rotation), ${\cal F}(0)=0.068$ (non-rotating black hole), and ${\cal F}(1)=0.5$ (extreme rotation). 

Let us compare the prediction of equation (\ref{M-J}) with the data \citep{mr} for the three binary black holes with 3/2 or 3/2/1 frequency ratios and measured mass. Using the lower limit ${\cal F}(a)>0.038$, we obtain $M>8.2M_\odot$ for GRO J1655--40 ($f=300, 450$ Hz), observed to have $M=6.3\pm 0.3 M_\odot$. Similarly for XTE J1550--564 ($f=92, 184, 276$ Hz), we obtain $M>13.3 M_\odot$, compared to the observed $M=9.6\pm 1.2 M_\odot$. There are two pairs of 3/2 QPOs observed in GRS 1915+105 ($f=41, 67; 113, 168$ Hz), whose black hole has a measured mass of $M=14\pm 4 M_\odot$. For the higher frequency pair, we obtain $M>22M_\odot$ while for the lower frequency pair we obtain $M>58M_\odot$. If we use no rotation rather than counter rotation to provide the lower mass limits, they would increase by a factor 1.8. Similar conclusions have been reached by \citet{tb}.

It is perhaps not surprising that these QPOs cannot be explained as $m\ge 1$ g--modes, since they occupy a much smaller region of the disk nearer to its inner edge (where our accretion disk model is most suspect)\citep{per}. In addition, their $\phi$ dependence reduces the observed modulation. 

One would thus wonder why the $m=0$ g--mode was not seen. It provides a relation $M/M_\odot=F_0(a)/f_0$ similar to equation (RD.I)\citep{wso}. Indeed, all but two of the QPOs in these sources could be fundamental g--modes (for some value of $a<1$), but of course only one in each source. The value of $a$ required for some of these is close to that estimated from the spectroscopic method [temperature and luminosity determine $r_i(a)$]. For instance, \citet{smn} obtain $a=0.65 - 0.75$ for GRO J1655-40, whereas identifying its 300 Hz QPO as an $m=0$ g--mode requires $a=0.9 - 1.0$. \citet{mc} obtain $a>0.98$ for GRS 1915+105, whereas identifying its 168 Hz QPO as an $m=0$ g--mode requires $a=0.8 - 1.0$.

From their MHD simulations, \citet{tv} claim that the $m=2,3,\ldots$ g--modes can grew to dominance over those of $m=0,1$ via the Rossby wave instability. However, this required a large concentration of magnetic field between the black hole and the accretion disk. In addition, their simulation neglected the vertical structure of the disk. We should note that \citet{abt} found no g--modes in their shearing--box MHD simulations of a limited radial region of an accretion disk. There were some indications of the generation of p--modes within the MRI--induced turbulence, however.

Very recently, \citet{rm2} reported results of ideal hydrodynamic (2D and 3D) and MHD (3D) global numerical simulations of accretion disks. As in \citet{tv}, the nonrotating black hole was represented by a modified Newtonian gravitational potential. The evolution was typically followed for about $10^2$ orbital periods of the inner disk. From power spectra at many radii, the $m=0$ g--mode was seen in the hydro simulations at the predicted frequency and radial extent. It was not seen in the MHD simulations. However, because of the induced MRI turbulence, it would not be expected to be seen if it was at the same amplitude as in the hydro simulations. Because of the limited range of $\phi$ (with periodic boundary conditions) and frequency in the 3D simulations, the higher $m$ modes could not have been seen.

\subsection{Perturbative effects of buoyancy and viscosity}

We need to say a few words about a non-zero buoyancy manifested by a Brunt-V\"ais\"al\"a frequency $N_z=N_z(r,z)>0$. This frequency is involved in all the equations only via the expression ($\omega^2-N_z^2$), so the corotation resonance equation (\ref{cordef}) becomes
\[ \omega^2-N_z^2=0 \; . \label{Ncor} \]
One might expect that the buoyancy would be small in realistic accretion disks, since the magneto-rotational instability produces strong turbulence \citep{hk}, which should locally homogenize the specific entropy. If so, $N_z$ should be treated as a perturbation in the above equation. Because of that, it leads only to a small and generally $z$--dependent change in the corotation resonance position. In fact, the corotation resonance point splits into two nearby ones,
\[
\rc^\pm\approx\rc\pm\frac{N_z(\rc,z)}{m}\,\left[\frac{d\Omega(\rc)}{dr}\right]^{-1}\; .
\]
Our arguments and results regarding the mode divergence apply to both of them. 

In the presence of viscosity which acts hydrodynamically (via the $\alpha$ model) and perturbatively, \citet{ort} found that for most (including these g--) modes, a viscous instability is induced. Such accretion disks are thus secularly unstable.
An effective viscosity (in particular, generated by the magneto-rotational instability) should be present in these (thin) accretion disks, but it is not known in what ways it acts like a hydrodynamic viscosity. 
To lowest order in its magnitude, the viscosity does not change the values of the eigenfrequencies, but introduces a small imaginary part in them. As usual, this imaginary part removes the divergences found above \citep{nw91}, since the corotation resonance equality (\ref{cordef}) no longer holds at any radius (a well-known effect of a complex pole near the real axis). However, the actual oscillation amplitude will usually still be largest at the corotation resonance radius as long as the mode amplitude is linear.  

One is naturally tempted to contemplate the outcome of the mode growth. Will enough nonlinearity be induced to lead to significant mode--mode coupling and related effects? One would like to extend the local resonance analysis of \citet{ak} to this problem. However, one must also investigate how much the spatial extent of the modes averages over the strong MRI--induced turbulence. Could it be sufficient to provide a stochastic driving force (producing mode excitation) within a time--independent `unperturbed' state?

\acknowledgments

This work was supported by NASA grant NAS 8-39225 to Gravity Probe B.

\end{document}